# Highly-Oriented Atomically Thin Ambipolar MoSe$_2$ Grown by Molecular Beam Epitaxy

Ming-Wei Chen[1,2], Dmitry Ovchinnikov[1,2], Sorin Lazar[3], Michele Pizzochero[4], Michael Brian Whitwick[1], Alessandro Surrente[5], Michał Baranowski[5,6], Oriol Lopez Sanchez[1,2], Philippe Gillet[4], Paulina Plochocka[5], Oleg V. Yazyev[4], Andras Kis[1,2*]

[1] Electrical Engineering Institute, École Polytechnique Fédérale de Lausanne (EPFL), CH-1015 Lausanne, Switzerland
[2] Institute of Materials Science and Engineering, École Polytechnique Fédérale de Lausanne (EPFL), CH-1015 Lausanne, Switzerland
[3] FEI Electron Optics, 5600 KA Eindhoven, The Netherlands
[4] Institute of Physics, École Polytechnique Fédérale de Lausanne (EPFL), CH-1015 Lausanne, Switzerland
[5] Laboratoire National des Champs Magnétiques Intenses CNRS-UGA-UPS-INSA, 143 avenue de Rangueil, Toulouse, France
[6] Department of Experimental Physics, Faculty of Fundamental Problems of Technology, Wrocław University of Science and Technology, Wybrzeze Wyspianskiego 27, 50-370 Wroclaw Poland

*E-mail address: andras.kis@epfl.ch

Transition metal dichalcogenides (TMDCs), together with other two-dimensional (2D) materials have attracted great interest due to the unique optical and electrical properties of atomically thin layers. In order to fulfill their potential, developing large-area growth and understanding the properties of TMDCs have become crucial. Here, we used molecular beam epitaxy (MBE) to grow atomically thin MoSe$_2$ on GaAs(111)B. No intermediate compounds were detected at the interface of as-grown films. Careful optimization of the growth temperature can result in the growth of highly aligned films with only two possible crystalline orientations due to broken inversion symmetry. As-grown films can be transferred onto insulating substrates allowing their optical and electrical properties to be probed. By using polymer electrolyte gating, we have achieved ambipolar transport in MBE-grown MoSe$_2$. The temperature-dependent transport characteristics can be explained by the 2D variable-range hopping (2D-VRH) model, indicating that the transport is strongly limited by the disorder in the film.

Atomically thin two-dimensional (2D) materials have shown great potential because of their interesting electrical and optical properties.[1–4] Potential applications in flexible electronics and the possibility to further extend their range of applications by integrating them into heterostructures[5] motivate scientists to develop a reliable way to grow large-area 2D materials.[6–9] While chemical-vapor-deposition (CVD) can on one hand result in high-quality 2D materials and 2D heterostructures with sharp and clean interfaces[10], the toxicity of some of the precursors introduces challenges, and the lack of in-situ characterization and monitoring during growth can lead to non-reproducible results. On the other hand, using molecular-beam-epitaxy (MBE) to grow atomically thin transition metal dichalcogenides (TMDCs) and other 2D materials via van der Waals epitaxy[11–16] has several potential advantages. The deposition can be well controlled using high-purity elemental sources, direct heterostructure growth can be achieved, and the growth in an ultra-high vacuum environment limits the amount of impurity atoms. In addition, the quality of the MBE-grown film can be monitored during growth in-situ using reflection-high-energy-electron-diffraction (RHEED). The growth of different TMDCs on various substrates has been recently demonstrated,[17–19] and the band structures and thin film morphology have been intensively investigated using angle-resolved photoemission spectroscopy and scanning tunneling microscopy.[20–22] However, reports on the electrical properties of atomically thin layers grown by MBE[23] are rare and indicate that the material quality needs to be improved further. Starting from this point, it is crucial to optimize epitaxial growth and to investigate the optical and electrical properties of MBE-grown atomically thin TMDCs.

Here, we report on the use of MBE to grow atomically thin MoSe$_2$ on GaAs(111)B down to nominal monolayer (ML) thickness, exhibiting high optical quality confirmed by photoluminescence (PL) and Raman spectroscopy. We find that using GaAs(111)B as the growth substrate results in a high degree of control over lattice orientation of the domains forming the polycrystalline film. The features of van der Waals epitaxy were thoroughly investigated by X-ray photoelectron spectroscopy (XPS). Using scanning transmission electron microscopy (STEM), we have observed that depending on the growth temperature, polycrystalline MoSe$_2$ films can be composed of grains with either a 15° or a 60° misorientation angle. Both configurations are predicted to be energetically stable by density functional theory (DFT) calculations. We further address the electrical properties of highly-oriented nominal ML MoSe$_2$, and demonstrate ambipolar transport behavior using polymer electrolyte gating.

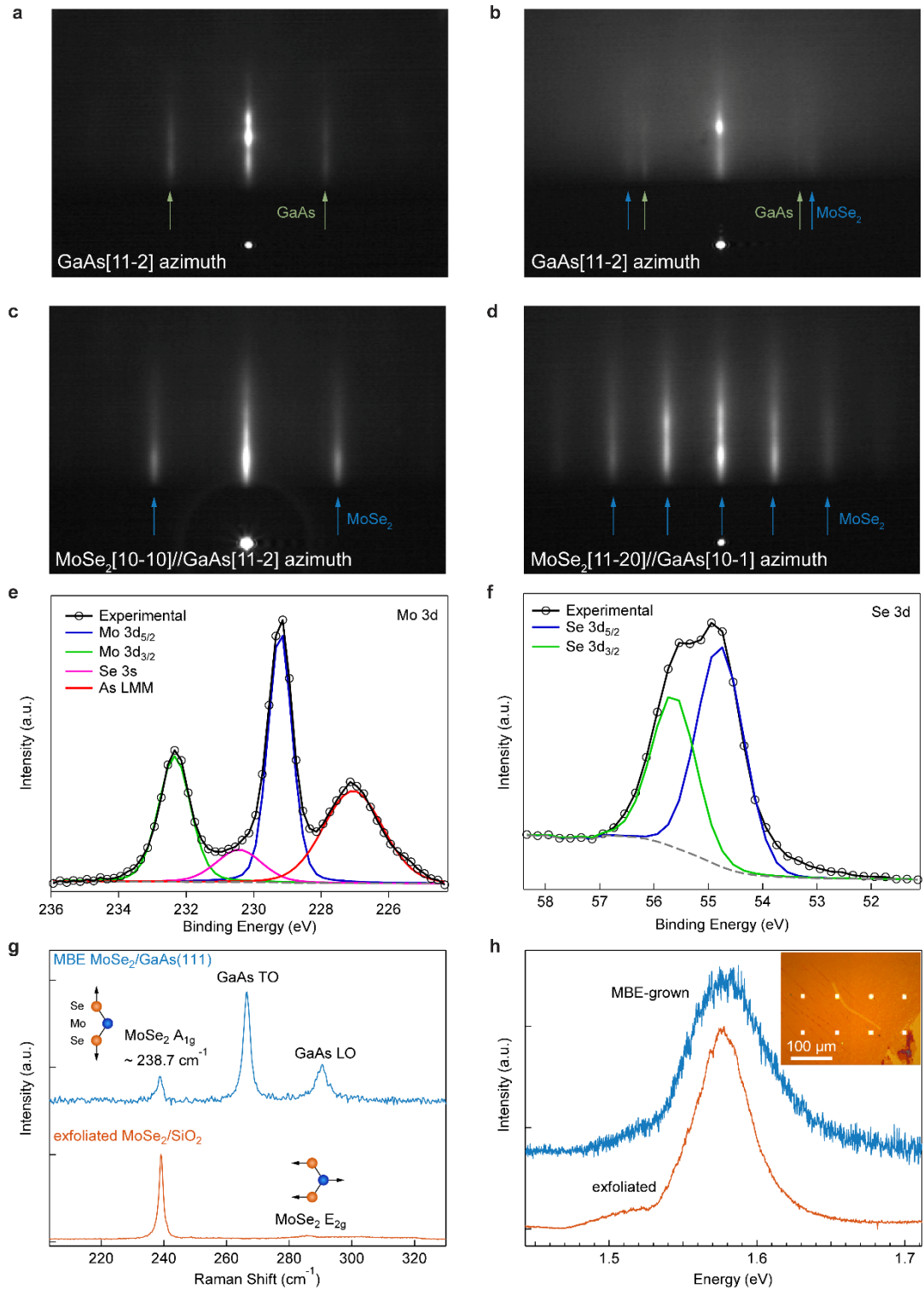

Figure 1. Growth of atomically thin MoSe$_2$ by MBE. (a) RHEED patterns of GaAs(111) along the GaAs[11-2] azimuth at growth start. (b) Half-half transition along the GaAs[11-2] azimuth. (c) (d) Nominal monolaer (ML) MoSe$_2$ observed along MoSe$_2$[10-10] and MoSe$_2$[11-20] azimuths at growth end after 22 min. (e) Mo 3d and (f) Se 3d core-level spectra in XPS. (g) Comparison of Raman spectra from MBE-grown and exfoliated ML MoSe$_2$. (h) Photoluminescence of transferred ML MoSe$_2$ and exfoliated ML MoSe$_2$. Inset shows optical image of the transferred film. The sample was grown at 470 °C.

Temperature-dependent measurements indicate that the transport is limited by the disorder and can be explained using the 2D variable range hopping (2D-VRH) model.

Atomically thin MoSe$_2$ films were grown on GaAs(111)B substrates by MBE and the growth was monitored *in-situ* using a RHEED camera. Atomic hydrogen was used to effectively remove native oxide at low temperature without As desorption,[24,25] and a clean GaAs(111)B surface with sharp RHEED streaks along the GaAs[11-2] azimuth was observed (see Figure 1a). As the growth progresses, GaAs-related streaks gradually fade out and are replaced by a newset of faint streaks with a smaller spacing, indicating the formation of MoSe$_2$. A RHEED pattern from the MoSe$_2$ film with a half-layer coverage is shown on Figure 1b, with GaAs and MoSe$_2$ streaks having similar intensities. GaAs-related streaks completely vanish as a full film is grown, while MoSe$_2$ streaks become more intense Figure 1c. The sharp streaks in Figure 1c and Figure 1d indicate the epitaxy with MoSe$_2$[10-10]//GaAs[11-2] and MoSe$_2$[11-20]//GaAs[10-1], respectively. The van der Waals epitaxy thus provides a reliable way for developing large-scale 2D materials on different substrates.[19,26]

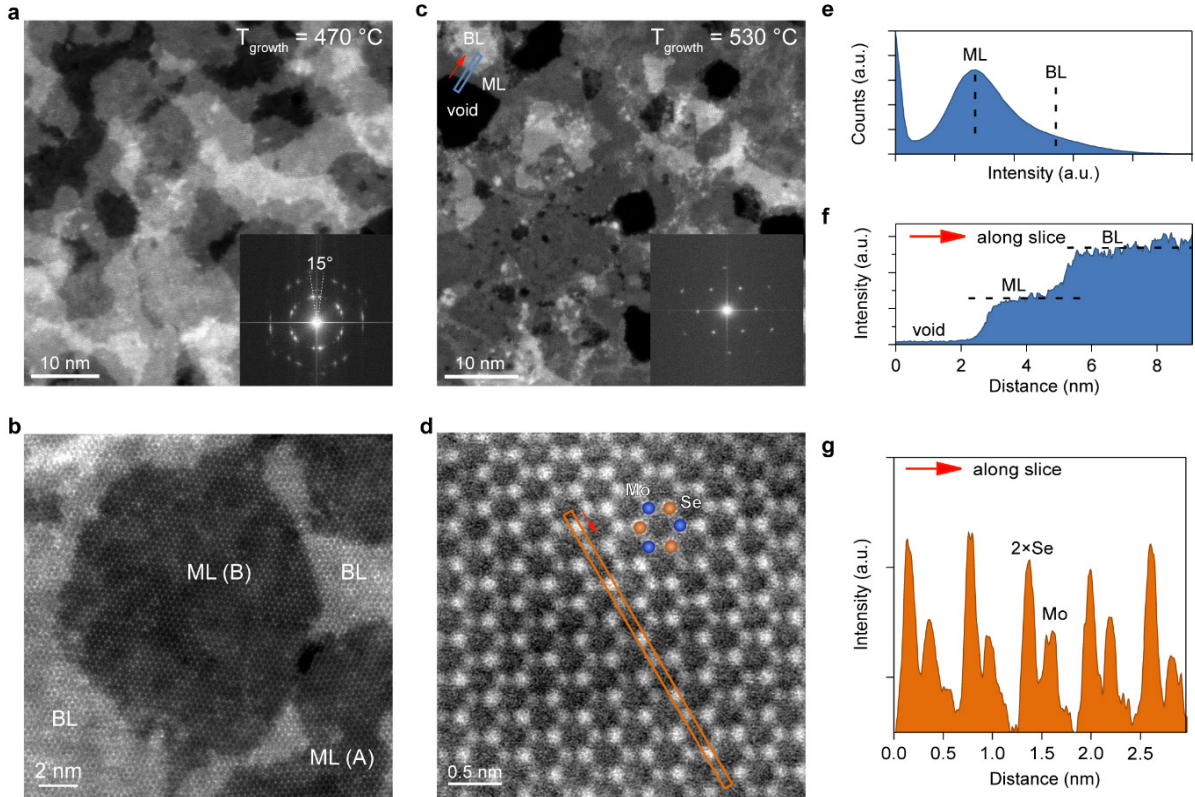

Figure 2. Morphology of MBE-grown MoSe$_2$ films. (a) Low-magnification HAADF-STEM image of MoSe$_2$ grown at 470°C. Inset is the corresponding FFT image showing two sets of spots. (b) High-magnification image of an incomplete bilayer (BL) comprised of ML domains with two orientations denoted as ML(A) and ML(B). (c) Low-magnification HAADF-STEM image of MoSe$_2$ grown at 530 °C. Inset shows the corresponding FFT image with a single set of diffraction patterns. (d) High-magnification of ML region with schematic of atom positions. Bright spots correspond to double Se atoms (e) Intensity histogram of the image shown in (c). (f) Intensity profile along the slice shown in (c). (g) Intensity profile of the slice from (d).

Films grown on GaAs(111)B were examined *ex-situ* using x-ray photoelectron spectroscopy (XPS). Figure 1e shows the core level spectrum of the Mo 3d range. The binding energies of Mo 3d$_{5/2}$ and Mo 3d$_{3/2}$ peaks are 229.2 eV and 232.4 eV, respectively. An additional As$_{LMM}$ peak with a lower binding energy of 227.2 eV is also observed. The Se 3d core level spectrum in Figure 1f shows a Se 3d$_{5/2}$ peak at 54.9 eV and an Se 3d$_{3/2}$ peak at 55.8 eV, demonstrating the existence of Mo-Se bonds. A consistent binding energy shift is observed in the core level spectra of Ga 3d and As 3d between pristine GaAs substrate and as-grown MoSe$_2$, implying that charge transfer takes place at the interface (see Supplementary Section 1 for more details). On the other hand, the peak positions and the high quality of fitting shows that no intermediate compounds exist at the interface, as expected from van der Waals epitaxy.

The Raman spectrum of nominal ML MoSe$_2$/GaAs(111) shown on Figure 1g clearly shows the MoSe$_2$ A$_{1g}$ mode at 238.7 cm$^{-1}$ which is comparable with that of exfoliated ML MoSe$_2$ on 270 nm SiO$_2$. Photoluminescence spectroscopy was also used to investigate the optical properties of ML MoSe$_2$. In order to avoid the optical quenching effect from the substrate,[18,27] the as-grown film was transferred onto a 270 nm SiO$_2$/Si chip with the optical image shown in the inset of Figure 1h and Figure S2. The 1.58 eV peak recorded at room temperature is clearly shown in Figure 1h and is comparable with the exfoliated ML flake in terms of the peak positon. Despite inevitable existence of wrinkles and folded regions due to the imperfect transfer process which might introduce defects and peak broadening, the full-width-half-maximum of 37 meV is comparable to that of exfoliated ML flakes (*FWHM* = 27 meV).

The films were transferred onto the TEM grids and investigated using Cs-corrected STEM in order to examine their morphology. Figure 2a shows a low-magnification high-angle annular dark-field (HAADF)-STEM image of MoSe$_2$ grown at the temperature of 470 °C. The patch-like islands in a few-nm scale with different brightness represent regions with varying thicknesses. Surprisingly, the corresponding fast Fourier transform (FFT) image calculated from this image and shown in the inset of Figure 2a shows only two sets of spots with a six-fold symmetry, rotated by ~15° with respect to each other. The high-magnification HAADF-STEM image in Figure 2b shows corresponding domains with a relative rotation of 15°, labeled ML (A) and ML (B). These are sometimes also stacked on top of each other, forming bilayers (BLs). A more detailed analysis of the grain orientation is presented on Figure S3 and shows that both orientations appear with roughly the same frequency in the ML film. Most of the BL area (~70%) is however of the ML(A) + ML(B) type, composed of two MoSe$_2$ layers with an interlayer twist of 15° and could present an interesting material system for studying the effect of interlayer twist on electrical and optical properties of 2D semiconductors.[28]

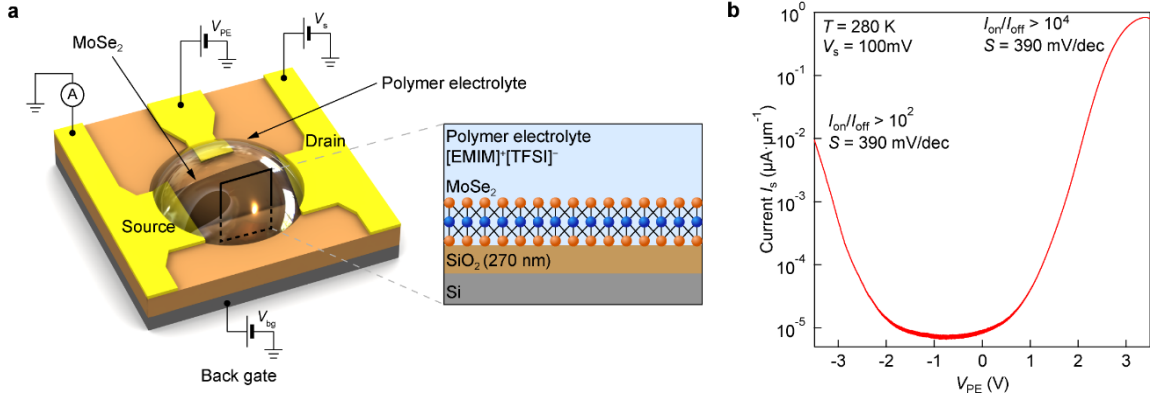

Figure 3. Ambipolar transport in MoSe$_2$ EDLT. (a) Schematic of the MoSe$_2$ EDLT in a dual-gate geometry. The back-gate is used to modulate the charge density in the 2D channel around the value set by the reference electrode with the applied voltage $V_{PE}$. (b) Channel current as a function of $V_{PE}$ showing ambipolar behavior.

We have gained further insight into the stability of both types of MoSe$_2$/GaAs (111)B superlattices by considering several models and determining their formation energies through density functional theory calculations (see Supplementary Section 3). The picture that emerges suggests that a twisting angle not only dramatically reduces the strain in the MoSe$_2$ lattice, but additionally leads to a slightly more favorable interaction with the GaAs (111)B substrate, irrespectively of the relative MoSe$_2$ − GaAs orientation. Overall, formation energies of oriented and misoriented MoSe$_2$/GaAs (111)B superlattices are very similar, indicating that both cases are likely to form. Nevertheless, it is also worth mentioning that the growth dynamics should also play a role with the migration of adatoms prior to in-plane bond formation which could influence the growth, resulting in large regions with single orientation. Possible changes in the surface reconstruction at these temperatures close to GaAs decomposition are not taken into account either, which could explain the difference in the morphology of films grown at these two temperatures.

A slight increase of growth temperature, into the 500 − 530 °C range, results in increased order in the film with the grains no longer showing the 15° misorientation. The HAADF-STEM image of the film is shown on Figure 2c. The corresponding FFT image now shows only one set of peaks. Since ML MoSe$_2$ does not possess inversion symmetry, we cannot exclude the presence of grains with a 60° relative orientation at this point, since these would result in a set of diffraction peaks at the same positions in Fourrier space. The presence of voids in the film indicates that BLs start to form before the first MLs complete. Similar morphologies of MBE-grown MoSe$_2$ on highly oriented pyrolytic graphite, graphene and SiC(0001) were also observed by STM by other groups.[18,21,27,29,30]

Since the intensities recorded in HAADF-STEM images are related to Rutherford scattering which increases with the atomic number (Z),[31,32] different intensities in the image can be attributed to different layer thickness. Figure 2b shows the intensity histogram with dashed lines that correspond to intensities in ML and BL regions. The intensity profile along the slice of interest is plotted in Figure 2c, showing thickness varying from ML to BL. Figure 2d shows a high-magnification image of a ML region with the intensity profile along a slice of interest shown in Figure 2f, where the positions of Mo and Se atoms can be assigned to periodically varying intensities. The ML has a 2H structure with a lattice constant estimated to be 3.29 ± 0.03 nm, which is in line with the bulk lattice constants reported in literature.[33,34]

To further confirm the long-range uniformity of the MoSe$_2$ film, we have measured second harmonic generation (SHG) from as-grown MoSe$_2$/GaAs(111)B and suspended MoSe$_2$ on TEM grids, Figure S7. Polar plots of SHG intensity show six-fold symmetry, while the PL intensity and Raman $A_{1g}$ peak position and intensity maps indicate a high degree of uniformity (see Supplementary Section 4).

We now focus on the electrical properties of 0, 60° MoSe$_2$ with nominal ML thickness. We use electrical double-layer transistors (EDLT) in order to access a wide range of electrostatically induced doping levels and to reduce Shottky barrier heights at contacts, allowing efficient electron and hole injection using the same contact material.[35] Figure 3 shows the schematic of the device. The polymer electrolyte PS-PMMA-PS:[EMIM]-[TFSI] is spin-coated on top and allows us to reach high carrier densities, increases the efficiency of carrier injection, making it ideal for achieving ambipolar regime of operation.[35,36] We also include a back gate, allowing charge carrier modulation in the semiconducting film at temperatures below the freezing point of the polymer electrolyte ($T$ ~ 200 K). Devices without a polymer electrolyte show a very poor transistor behavior (see Supplementary Section 5). The dependence of the EDLT on the polymer electrolyte voltage $V_{PE}$ exhibits a clear ambipolar behavior close to room temperature ($T$ = 280 K) (Figure 3b). We find a current $I_{on}/I_{off}$ ratio of ~10$^4$ and ~10$^2$ for the $n$ and $p$ sides, respectively. The maximum current density on the $n$-side, with the value of ~1 µA/µm is two orders of magnitude larger than on the $p$-side, possibly due to intervalley scattering of electrons, while the off current remains at pA levels. The subthreshold swing calculated from the linear region is for both sides ~390 mV/dec. Field-effect mobilities can be extracted from four-contact devices by freezing the polymer electrolyte at 200 K and performing a back-gate sweep (see Supplementary Section 6) which allows the charge carrier concentration in the 2D semiconductor to be modulated around the value set by $V_{PE}$ prior to freezing the electrolyte. The extracted electron mobility $\mu_e$ is ~ 0.05 cm$^2$ V$^{-1}$ s$^{-1}$ and hole mobility $\mu_h$ is ~ 0.28 (cm$^2$ V$^{-1}$ s$^{-1}$). The mobility values are significantly lower than those of CVD-grown MoSe$_2$.[37,38] These results indicate that charge carrier transport is strongly influenced by the disorder in the film.[23,39]

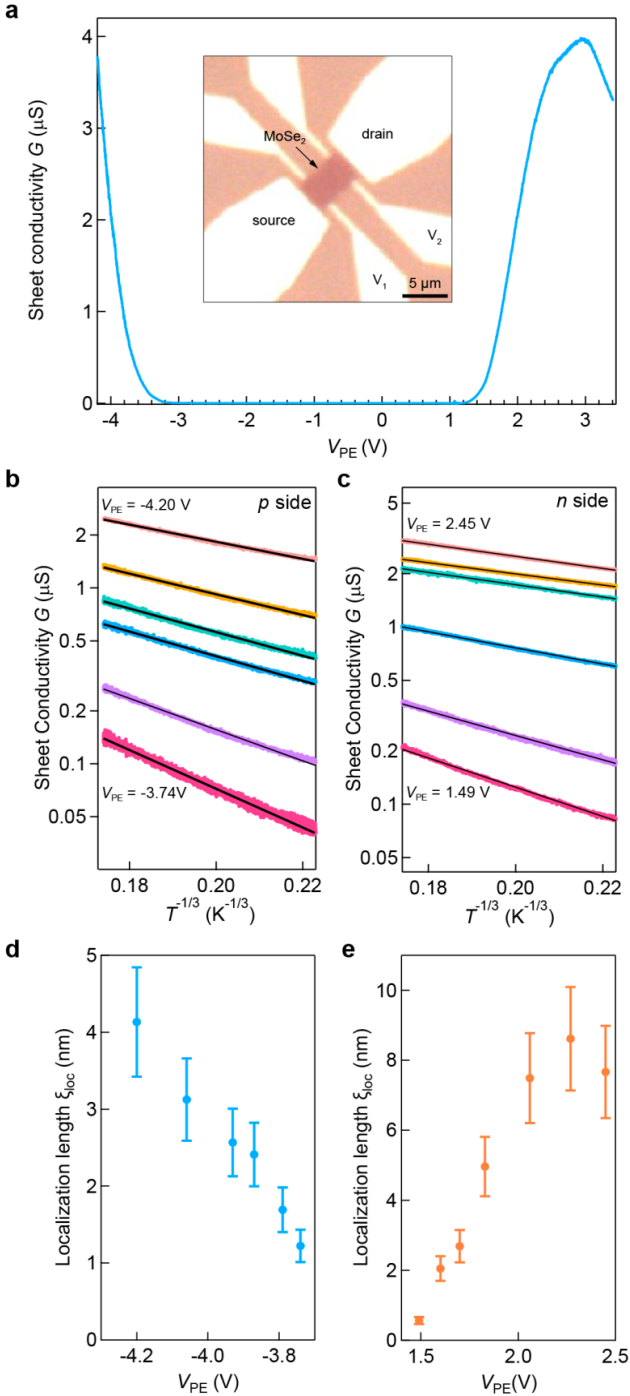

Figure 4. 2D-VRH transport mechanism in MoSe$_2$ EDLT. (a) Sheet conductivity $G_{sh}$ as a function of $V_{PE}$ at 280 K. Inset: optical image of the four-contact device. (b) and (c) $G_{sh}$ as a function of $T^{-1/3}$ on the hole and electron sides for different values of $V_{PE}$ filled lines correspond to fits to the VRH model. (d) Evolution of the localisation length $\xi_{loc}$ with $V_{PE}$, extracted from fits to the VRH model on the hole branch. (e) Dependence of $\xi_{loc}$ on $V_{PE}$ for the electron branch.

In order to elucidate the dominant transport mechanism in MBE-grown atomically thin MoSe$_2$, we have performed electrical measurements as a function of temperature, polymer electrolyte and back-gate voltage. Figure 4a shows the sheet conductivity $G_{sh}$ as a function of $V_{PE}$ based on the four-contact device shown in the inset. The $V_{PE}$ applied for p-side needs to be pushed to $V_{PE} > -4V$ to reach the same value of $G_{sh}$ of n-side, indicating strong electron doping or Fermi level pinning to the conduction band. A drop in $G_{sh}$ takes place at $V_{PE} > 3V$ and can be attributed to the electrolyte-induced disorder which is commonly observed in experiments involving EDLTs.[40,41] The sweep at 280 K can then provide a reference curve for different doping levels that can be achieved by changing $V_{PE}$. Once the doping level at a given $V_{PE}$ is stabilized at 280 K, we freeze the electrolyte down to 200K with a cool-down rate of 1 °C/min. The $V_{PE}$ is then disconnected at 200 K after the electrolyte was completely frozen so that $G_{sh}$ can be stabilized (see Supplementary Section 7). The $G_{sh}$ during each cooldown was recorded down to 12 K with a cooling rate of 0.5 °C/min. The process was reversible after a mild annealing by ramping the temperature to 333 K.

The $G_{sh}$ of MoSe$_2$ at a given $V_{PE}$ monotonically decreases with decreasing temperature, showing semiconducting behavior for both sides. The dependence above 80 K follows the 2D-VRH model[42] with the relation $G_{sh} \propto \exp[-(T_0/T)^{1/3}]$ where $T_0$ is the characteristic temperature. A linear fit of ln $G_{sh}$ to $T^{-1/3}$ is plotted at each given $V_{PE}$, demonstrating the validity of the 2D-VRH mechanism for both sides, Figure 4b and c. The VRH mechanism is usually observed in disordered systems[23,40] and the results imply that the transport of MoSe$_2$ EDLT is strongly influenced by the voids and the nanometer-scale grains in the film.

Charge carriers are strongly scattered and tend to hop between different conductive paths. The localization length can be changed in a small range by changing $V_{PE}$, i.e., changing the doping level. The dependency is evident by extracting $T_0$ from the fits to the 2D-VRH model with the value of slopes $s$ extracted from Figure 4b and Figure 4c, where $T_0 = s^3$. The values of $T_0$ decrease by more than one order of magnitude with the increase of $|V_{PE}|$ because of the increase of carrier densities in the material, thus screening the disorder along the conductive paths. The localization length $\xi_{loc}$ can be extracted using the expression $\xi_{loc} = \sqrt{13.8/k_B D T_0}$, where $k_B$ is the Boltzmann constant and $D$ the density of states (see Supplementary Section 8). The results are plotted in Figure 4d and e for p and n sides, respectively. Holes have a slightly lower $\xi_{loc}$ with $\xi_{loc}$ reaching a maximum value of ~ 4 nm at $V_{PE} = -4.2$ V. The n-side on the other hand shows tunable $\xi_{loc}$ up to ~ 9 nm with $V_{PE} = 2.6$ V. All values of $\xi_{loc}$ have the same order of magnitude as the grain size shown in STEM images, indicating that the 2D-VRH transport is linked to the disorder in the as-grown MoSe$_2$. Future MBE-based growth efforts will have to concentrate on increasing the grain size in order to improve the film quality.

In conclusion, we have grown atomically thin MoSe$_2$ using MBE. Films show a high degree of alignment due to the van der Waals interaction with the GaAs substrate and can be transferred to insulating substrates for further optical and electrical transport studies. We realize electrolytically gated transistors based on transferred ML MoSe$_2$ films. Electrical transport follows 2D-VRH model due to the disorder in film, with localization length comparable to the grain size.

## METHODS

### MBE Setup and Material Growth

The growth was carried out in an Omicron MBE (Lab 10) with a ~$10^{-10}$ mBar base pressure. Cleaved 1×1 cm$^2$ GaAs(111)B substrates were outgassed at up to 500 °C for at least 30 min. The native oxide was removed from the surface of GaAs(111)B by heating it to 350 °C under a flux of atomic hydrogen. Hydrogen molecules were dissociated by a tungsten filament with Joule heating at 70 W and were introduced into the chamber *via* a leak valve. The procedure lasted 30 min or more at base pressure ~$3\times10^{-7}$ mBar resulting in sharp streaks in RHEED. A Kundsen cell and an electron beam source (EFM-3 from Omicron) were used for Se and Mo evaporation, respectively. The flux rates were calibrated using a quartz crystal microbalance, and the flux ratio of Se/Mo was optimized to be ~ 40 for growth. A RHEED camera (Staib Co.) was used to monitor the growth *in-situ*. The growth temperature was optimized in the 470 − 530 °C temperature range. Post annealing at up to 550 °C was performed in Se atmosphere. Higher temperature leads to GaAs decomposition and increasing surface roughness.

### XPS and Raman Spectroscopy

The XPS spectra were obtained *ex-situ* in a commercial KRATOS AXIS ULTRA system, and C1s core level peak at 284.8 eV was used for the reference. Peak identification and fitting was performed in PHI MultiPak processing software. Raman analysis was performed using a Horiba LabRAM HR800 system using a 532-nm-wavelength green laser with spot size ~ 4 μm. The laser power was kept below 4 mW during all measurements. We used a 1800 lines/mm grating and have calibrated the system using the polycrystalline Si peak at 520 cm$^{-1}$. The PL was measured in a home-built setup using a 488 nm laser (Coherent) for excitation.

### STEM Microscopy and Analysis

The STEM experiments were performed on a FEI Titan Themis 300 double Cs corrected microscope at an acceleration voltage of 80 kV in order to minimize beam damage. The scanning probe had a 28 mrad semiconvergence angle allowing resulting in a resolution close to 1 Å. The data was acquired under annular dark field conditions using an annular detector with a collection half-angle between 40 and 200 mrad.

### Material Transfer and Device Fabrication

As-grown films were coated with PMMA and immersed into 30% KOH$_{(aq)}$ at 90 °C. The detached PMMA layer with the as-grown film was then transferred to a beaker with deionized water several times to remove excess KOH$_{(aq)}$ and was transferred onto a degenerately doped n$^{++}$ Si chip covered by 270 nm SiO$_2$ or TEM grids. PMMA was removed in acetone. Palladium was used for electrical contacts in a standard PMMA-based e-beam lithography process. A second e-beam lithography was performed, followed by O$_2$/SF$_6$ plasma etching in order to define the device geometry. To fabricate an EDLT, the electrolyte PS-PMMA-PS:[EMIM]-[TFSI] was spin-coated onto the device and soft-baked at 60 °C for 10 min. A more detailed description is available elsewhere.[43,44] Electrical measurements were carried out using an Agilent 5270B SMU and Keithley 2000 DMM. Cryogenic measurements were performed in a Janis closed-cycle cryogen-free cryostat.

## ACKNOWLEDGEMENTS

We thank R. Gaal for technical assistance with the Raman setup, D. Alexander (CIME) for support with electron microscopy and P. Mettraux for help with the XPS setup and experiments. We thank D. Dumcenco, H. Kim, and A. Pulkin for fruitful discussions. This work was financially supported by the European Research Council grants nos. 240076 and 306504, the Swiss National Science Foundation grant nos. 153298 and 162612, funding from the European Union's Seventh Framework Programme FP7/2007-2013 under Grant Agreement No. BLAPHENE project under IDEX program Emergence and Programme Investissements d'Avenir under the program ANR-11-IDEX-0002-02, reference ANR-10-LABX-0037-NEXT. This work was financially supported by funding from the European Union's Seventh Framework Programme FP7/2007-2013 under Grant Agreement No. 318804 (SNM.) and was carried out in frames of the Marie Curie ITN network "MoWSeS" (grant no. 317451). We acknowledge funding by the EC under the Graphene Flagship (grant agreement no. 604391). First-principles calculations were performed at the Swiss National Supercomputing Centre (CSCS) under the project s-675.

### Supporting Information Available

Supplementary figures and discussion related to XPS spectroscopy, TEM imaging, DFT calculations, SHG and Raman mapping, field-effect transistor characterization and variable range hopping transport mechanism. This material is available free of charge *via* the Internet at http://pubs.acs.org.